# Cavity enhanced terahertz fingerprint detection with ultrahigh sensitivity

Xiaomei Shi, and Zhanghua Han*

*Lab of Terahertz Photonics, China Jiliang University, Hangzhou 310018, China*

*Corresponding author: han@cjlu.edu.cn*

We report a new scheme of realizing terahertz fingerprint detection with ultrahigh sensitivity. Instead of using the direct absorption of terahertz through a bare sample in the regular transmission scheme, a cavity mode resonating at the characteristic frequency of the sample is used and due to the high dependence of the cavity mode transmission on the material loss, an amplified transmission decaying is observed when the sample is loaded into the cavity. Furthermore, this scheme retains the feature of substance identification. A one-dimensional photonic crystal cavity is used as the example for the detection of α-lactose and an efficient detection of 7nm α-lactose can be achieved, which corresponds to 1/80000 of the free space wavelength at the characteristic frequency of 0.529THz, exhibiting sensitivity 500 times higher than the regular method.

Terahertz (abbreviated as THz and typically defined between 0.1THz and 10THz) radiations have many advantages over the electromagnetic waves in other frequencies due to their favorable properties like transparency in most dielectrics, lower photon energy and nonionizing features, which may promise many potential applications ranging from fundamental sciences to practical applications [1]. Especially, many chemicals and molecules have their characteristic absorption frequencies located in terahertz regime resulting from their rotation, intra- and inter-molecular vibrations, suggesting that terahertz technology can work as a unique tool for chemical identification [2]. This technique, known as fingerprint detection, lays the foundation for many applications of terahertz in security checking (like drugs and explosives), biomedical diagnosis and pharmaceutic industry. However, due to the high contrast between the nanometer-scale size of most molecules and terahertz wavelengths (from tens to a few hundreds of microns), the interactions between molecules and THz radiations are extremely weak and a large volume of chemicals are usually required to have an observable absorption for identification. For example, in the mostly-used transmission scheme for THz fingerprint detection, the sample is normally made into powder and then compressed into pellets with a thickness of several millimeters [3]. There are many circumstances in which the sample is not abundant or it is required to use as few sample as possible, e.g. in medical diagnosis. Thus, a substantial improvement of the sensitivity in the THz fingerprint detection is still necessary to further broaden this technique in practical applications. Many approaches have been attempted in this respect, for example to use resonating antennas, metamaterials or InSb-based plasmonic gratings [4-7] in the transmission mode to enhance the local electric field so that the absorption cross section of the chemicals can be amplified. Other examples include the exploration of new working principles like hybridization induced transparency [8] or using waveguiding structures [9] to increase the interaction length between molecules and terahertz radiations for a fixed amount of sample. However, with all these reported approaches, to get an observable dependence of the transmission at the characteristic frequencies on the sample amount, the required sample thickness is on the order of several micrometers [6, 7]. In this paper, we propose a novel idea of realizing THz fingerprint detection with an ultrahigh sensitivity and the sample as thin as a few tens of nanometers can be easily and steadily captured.

Basically, to realize the functionality of fingerprint detection in the regular transmission scheme using a THz spectroscopy, one needs the information for two features. One is the characteristic frequencies which are the spectral signatures of a certain sample, manifesting them as resonances in the transmission spectrum; the other is the change of the transmittance at these frequencies (usually normalized using the transmittance through the sample to that through vacuum) which implies the amount of samples included in the measurement.

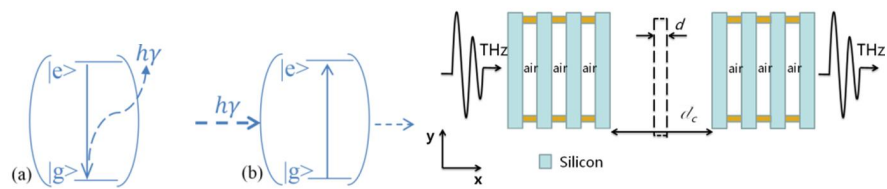

Fig. 1 Schematics for (a) the change of spontaneous rate of an emitter placed inside a cavity in the weak coupling regime; (b) the amplified transmission drop of a cavity mode with the frequency matching the characteristic frequency of the sample loaded into the cavity and (c) the 1D photonic crystal cavity compose of one central defect layer between two distributed Bragg reflectors made from four silicon stacks. The yellow parts are double-sided adhesive used to connect different parts while the dashed area indicates the loading of sample.

It is well-known in the weak-coupling regime that when an emitter is placed into a man-made cavity structure which has a designed resonance frequency corresponding to the energy difference between the ground and excited states of the emitter, as shown in Fig. 1(a), the spontaneous emission rate of the emitter can be enhanced due to Purcell effect [10]. Inspired by this picture, we conceived an idea of using a similar structure for THz fingerprint detection with ultrahigh sensitivity. The detection scheme is schematically shown in Fig. 1(b). A cavity structure is designed to have a resonating frequency $f_0$ which matches one of the characteristic frequencies of the target sample. When the sample is absent, the transmission through the cavity structure will exhibit a peak at $f_0$. If a thin target sample is present, due to the intrinsic absorption of the sample at $f_0$, the transmission will experience a drop and the resonance frequency will remain roughly at f0 due to the little optical path length change from the thin sample. This drop, however, will be much higher than the transmission through a bare sample due to the cavity effect. When another sample with an absorption frequency away from $f_0$ is considered, it works as a dielectric with small loss around $f_0$. So, when it is loaded into the cavity, the resonance will be spectrally shifted but its transmission won't be affected. One can see that using this scheme the sample can be identified while its amount can be measured with high sensitivity as well. The high dependence of the cavity mode transmission on the sample amount makes it possible to realize the quantitative detection easily.

In the following part, we demonstrate the capability of the proposed scheme for THz fingerprint detection using a one-dimensional (1D) photonic crystal (PC) cavity structures as an example. Composed of Si stacks, these structures have been experimentally investigated by us and the measured transmission spectrums demonstrate cavity resonances with high quality factors [11]. As shown schematically in Fig. 1(c), the structure consists of one central defect cavity between two parallel Bragg reflectors distributed on both sides. Each Bragg mirror is composed of four silicon stacks separated by air. Defect mode in the photonic bandgap can be formed due to the Fabry-Perot reflections inside the cavity. The refractive index of silicon and the air layers are 3.44 and 1, respectively. The length of the defect cavity ($d_c$) determines the numbers and frequencies of defect modes in this 1D photonic crystal cavity [11]. For simplicity, we selected a length of this defect cavity ($d_c$) being 319μm, which combing with other parameter, results in a transmission peak around 0.529 THz. This frequency is a signature of α-lactose, which will be used as the sample representative for the fingerprint detection in this paper. The thickness of the silicon layers is $t$ = 100 μm and air layer thickness between them $w$= 233μm.

The transmission characteristics of this photonic crystal cavity are numerically simulated with the finite element method (FEM). In the calculations, the incident THz plane wave is set be normal to the structure with electric field along the y direction. The calculated transmission spectrum of the bare photonic crystal cavity is shown in Fig. 2(a) which clearly illustrates the existence of the cavity mode at 0.529THz. An enlarged view of the transmission peak can be seen as the black line in Fig. 2(b), which shows that the peak is very sharp with a full-width at half-maximum bandwidth of 0.435 GHz, and the transmittance at resonance through this structure can approach 100%.

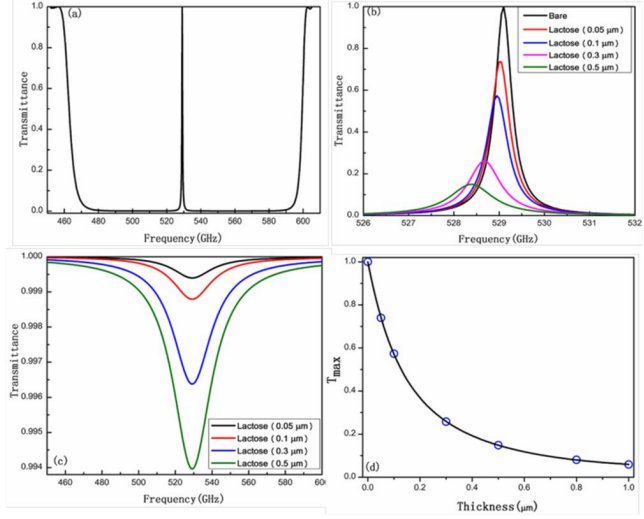

Fig. 2. (a) Simulated transmission spectrum of bare photonic crystal cavity with dc = 319 μm; t = 100 μm; w = 233 μm; (b) Transmission of 1D photonic crystal cavity with different thickness of α-lactose loaded at the center of this cavity. (c) The transmission through a bare α-lactose layer with different thicknesses. (d) Dependence of the transmittance through the cavity at resonance on α-lactose thickness.

When a thin layer of α-lactose is loaded into the center of the cavity, the peak transmission will be significantly affected by the intrinsic loss of it. The permittivity of α-lactose is modelled using with a series of Lorentzian oscillators as follows [12]:

$$\varepsilon_r = \varepsilon_\infty + \sum_{p=1}^{\infty} \frac{\Delta\varepsilon_p \omega_p^2}{\omega_p^2 - \omega^2 - j\gamma_p\omega} \qquad (1)$$

where $\varepsilon_\infty$ denotes the off-resonance background permittivity of α-lactose, $\omega_p$ and $\gamma_p$ are the angular frequency and damping rate of each absorption oscillation respectively and $\Delta\varepsilon_p$ is the oscillation strength factor. For simplicity only the first order absorption resonance of α-lactose at 0.529THz is considered and the other parameters are as follows $\varepsilon_\infty = 3.145$, $\gamma_p = 1.59 \times 10^{11}$ s$^{-1}$ and $\Delta\varepsilon_p = 0.052$, which together gives a calculated permittivity close to the empirical values [13]. When the α-lactose layer as thin as 0.05 μm is loaded at the center of the cavity, one can see from the red line in Fig. 2(b) that the transmittance at resonance drops by 26% dramatically from around 100% without lactose to be 74% while the position of the peak red-shifts remains almost unchanged. This is due to a significant dependence of the cavity mode on the material loss inside the cavity at resonance and a negligible change of the overall optical path due to the small thickness of α-lactose. As can be seen in Fig. 2(b), a successive increase of the α-lactose thickness will result in a further decrease of the transmittance at resonance, followed by a larger resonance shift due to a larger change in the optical path. The overall transmittance at resonance as a function of the α-lactose thickness is shown in Fig. 2(d) which exhibits an exponential decaying behavior. One may also notice that the resonance positions in Fig. 2(b) for different α-lactose thicknesses follow the trend of the black line which is the original transmission resonance of the bare PC cavity. This implies, to get a higher sensitivity of the transmittance drop as a function of α-lactose thickness, one should design an original cavity resonance with a higher quality factor. Actually, with the current cavity structure, if one assumes 5% is the limit of transmittance drop that can be steadily observed, our calculations show that value corresponds to a α-lactose thickness as small as 7 nm, which characterizes the ultimate resolution of our approach. One can also see in Fig. 2(d) that the resonance transmittance drop saturates at a α-lactose thickness around 1 μm. This dynamic range is determined by the thickness when the resonance peak shifts beyond the original cavity resonance in Fig. 2(b). In this aspect, to get a higher dynamic range, one should use a PC cavity resonance with a lower quality factor. A trade-off between the sensitivity and the dynamic range should always be balanced and this can be adjusted by changing the quality factor of the cavity mode.

For comparison, we also plot in Fig. 2(c) the transmission through such thin-layers of bare α-lactose. It is obvious that at resonance the drop in transmittance for a α-lactose thickness of 0.05 μm is only 0.06% and the number only increases to 0.34% and 0.59% even if the α-lactose thickness is 0.3 μm and 0.5 μm respectively. If one compares the numbers, it is found that the addition of the 0.05 μm α-lactose in the 1D photonic crystal cavity will introduce a drop in transmittance at the characteristic frequency which is about 433 times of that in a bare α-lactose layer of the same thickness. This number increases to roughly 500 at the α-lactose thickness of 7nm. In other words, the intrinsic absorption of α-lactose can be amplified by 500 times using our scheme. More importantly and worth noting, these transmittance drops indicated in Fig. 2(c) are probably not accessible with most terahertz spectroscopies due to the signal-to-noise issue while the drops in Fig. 2(b) can be steadily measured.

To further demonstrate that our scheme keeps the feature of substance identification similar to the conventional transmission scheme, we replace the α-lactose with another sample which has a different intrinsic absorption frequency. A new value of 2π*0.62 THz is used for $\omega_p$ in equation (1) while all the other parameters are assumed unchanged. The thickness of the new sample is 0.1 μm and the results of transmission spectra for three structures, the bare PC cavity, the PC cavity with 0.1 μm of α-lactose, and the PC cavity with 0.1 μm of the new sample with absorption at 0.62 THz, are shown in Fig. 3 for comparison. It can be seen that the transmittance for the PC cavity designed to identify the existence of α-lactose, only decreases when 0.1 μm of α-lactose is loaded while for the new sample the resonance of the PC cavity only experiences a red shift but with no decrease in the transmittance at all. This is because the new sample with an absorption frequency away from that of α-lactose will only introduce an additional optical path around 0.529THz in the cavity.

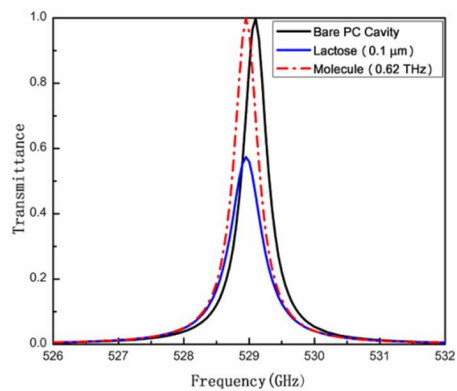

Fig. 3. A comparison of the transmission spectrum for three different cases: the bare PC cavity, the PC cavity with 0.1μm α-lactose and the PC cavity with 0.1μm of another sample with the absorption frequency at 0.62 THz.

Although a simple structure of 1D PC cavity is used to demonstrate the working principle of our proposed idea to realize terahertz fingerprint detection with unpresented sensitivity, we need to emphasize that the idea can be extended easily to other cavity structures working in the terahertz frequencies, like Si based waveguide ring resonators or photonic crystal cavities. Some on-chip cavity structures can even add more assets because the interaction length between the cavity mode and the sample is further increased.

In a summary, we have presented in this paper a new scheme to realize the THz fingerprint detection with ultrahigh sensitivity. Using a resonance mode matching the absorption of $\alpha$-lactose in a 1D PC cavity to amplify the influence of the $\alpha$-lactose intrinsic absorption to the change of transmittance at the resonance peak, a drop of more than a few hundred times can be obtained compared to that through a bare $\alpha$-lactose layer. With this method, an efficient detection of a thin $\alpha$-lactose layer with a thickness as small as 7 nm can be easily and steadily achieved. This thickness corresponds to only 1/80000 of the free space wavelength at the characteristic frequency of 0.529THz. A steady identification of such a thin layer of sample makes it possible to realize the recognition of a monolayer of molecules, suggesting huge potential for terahertz application in medical diagnosis. The proposed design can be easily extended to other substances by simply adjusting the cavity parameters to match the absorption frequencies of new materials. This versatility, combined with the ultra-high sensitivity, will make a significant impact on the current terahertz technology. It lays out the foundation for terahertz sensing in circumstance where high sensitivity is essential while still retaining the unique property of characteristic frequency identification. We believe that this technique really broadens the application range for the terahertz technology and paves the way for the further application of terahertz spectroscopy in biomedical area.